\documentstyle[preprint,aps]{revtex}
\tightenlines

\begin{document}
\title{Resonance tunneling of polaritons in 1-D chain with a single defect}
\author{Lev I. Deych and Alexander A. Lisyansky}
\address{Department of Physics,\\
Queens College of City University of New York, Flushing, NY 11367}
\date{\today}
\maketitle

\begin{abstract}
We consider propagation of coupled waves (polaritons) formed by a scalar
electromagnetic wave and excitations of a finite one dimensional chain of
dipoles. It is shown that a microscopic defect (an impurity dipole) embedded
in the chain causes resonance tunneling of the electromagnetic wave with the
frequency within the forbidden band between two polariton branches. We
demonstrate that resonance tunneling occurs due to local polariton states
caused by the defect.
\end{abstract}




Recently, there has been increasing interest in a novel kind of local states
that occur with frequencies inside a gap between different polariton
branches (``restrahlen'' region) when a dipole active defect is embedded in
a regular ionic crystal. These states are coupled states of transverse
electromagnetic waves and excitations of a crystal such as phonons or
excitons with both components localized in the vicinity of the defect.
Rupasov and Singh \cite{Rupasov} considered states of a two-level atom
coupled with elementary excitations of a medium. A two-level atom considered
in that paper could be called an ``active'' defect since it introduces new
modes (transitions between levels of the atom), which were absent in the
pure system. The local states associated with such a defect are the original
states of an atom modified by the interaction with its surroundings.
Correspondingly the frequency of the arising local state is the initial
frequency of an internal mode of the defect renormalized by the interaction
with the medium.

In our recent papers \cite{Deych,Podolsky} we considered a qualitatively
different situation. We showed that a defect ion with no internal degrees of
freedom embedded in a regular ionic crystal can give rise to local polariton
states. This kind of defects can be called a ``passive'' defect, and local
states predicted in Refs.\cite{Deych,Podolsky} occur due to fundamental
reconstruction of the spectrum of a pure system. As a result of such a
reconstruction, a discrete eigenfrequency splits from the continuous
spectrum giving rise to a local state.

This phenomenon is well known in systems of phonons or excitons and was
originally discovered by Lifshitz \cite{Lifshitz} (see also Refs.\thinspace 
\cite{Maradudin,Kosevich}). Similar effects were also found for
electromagnetic waves in photonic crystals \cite
{Yablonovich,Joannopulos,Figotin} due to macroscopic defects embedded in
their structure. Despite all the differences between local phonon states and
local photon states in photonic crystals, they share many common features.
For example, local states in both systems could appear in a 3-D system only
if the difference between the defect parameters (like a mass or an elastic
constant) and corresponding parameters of elements forming the pure
structure exceeds a certain threshold. The local polariton states are
essentially different from the states considered in Ref.\thinspace \cite
{Rupasov} as well as from local phonon states and from photon states in
photonic crystals. The most important general result obtained in
Refs.\thinspace \cite{Deych,Podolsky} is that a microscopic passive defect
is able to rebuild the spectrum of electromagnetic excitations in the medium
in the region of wavelengths much greater than the size of the defect.
Another striking difference between these local polaritons and other kinds
of local states arising due to passive defects is the absence of a threshold
for such states in isotropic systems even in 3-D situation. (For obvious
reasons there is no threshold for local states due to ``active'' defects in
any dimensions.)

In the present paper, we show that a dipole-active defect without additional
degrees of freedom embedded in an otherwise ideal structure causes resonance
tunneling of electromagnetic waves through the forbidden band. A similar
effect have been observed in a thin slab of a photonic crystal with a defect 
\cite{Yablonovich}. We would like to emphasize, however, that we discuss
resonance tunneling of electromagnetic waves with wavelengths much greater
than the characteristic scale of the defect, while in the case of photonic
crystals one deals with the situation when the wavelength is comparable with
the size of a macroscopic distortion of the periodicity. We show that the
resonance transmission in the stop-band occurs due to local polariton states
associated with the defect. Experimental observation of this effect seems
very promising and could be an unambiguous indication for local polaritons.

Another question that we would like to address is related to the fact that
local polaritons arise as a superposition of polariton modes from the entire
Brillouin band. Excitations with wave numbers at the edge of the band are
pure phonons and electromagnetic waves. In order to make sure that
short-wavelength components of the electromagnetic field do not destroy the
effect one must consider a model feasible for the microscopic treatment of
the field. In order to shed light on principle aspects of the problems
outlined above, we study light propagation through a one-dimensional finite
chain of dipoles with the nearest neighbor interaction. We assume that this
chain is placed within a single mode waveguide. Polaritons in this system
arise as coupled states of collective excitations of dipoles (polarization
waves) and electromagnetic waves. An important feature of the model is
accounting for the interaction between dipoles at different sites that leads
to spatial dispersion of the polarization waves. Taking into account the
spatial dispersion makes the exact analytical consideration of the problem
unfeasible even in the case of a pure system due to cumbersome algebra.
Therefore, we carry out numerical simulations by means of the
transfer-matrix method. The model can be described by the following
equations written in the frequency domain: 
\begin{equation}
(\Omega _{n}^{2}-\omega ^{2})P_{n}+\Phi (P_{n+1}+P_{n-1})=\alpha E(x_{n}),
\label{exciton}
\end{equation}
\begin{equation}
\frac{\omega ^{2}}{c^{2}}E(x)+\frac{d^{2}E}{dx^{2}}=-4\pi \frac{\omega ^{2}}{%
c^{2}}\sum_{n}P_{n}\delta (na-x),  \label{Maxwell}
\end{equation}
where the first equation describes the dynamics of site dipole moments, $%
P_{n},$ and the second one is the equation for the electric field $E$. Here $%
\Omega _{n}$ is the diagonal part of the force matrix responsible for the
short-range interaction between dipoles, and $\Phi $ is its off-diagonal
component. We assume that the defect, which occupies the site $n_{0}$
affects only the diagonal part of the force matrix, so that $\Omega
_{n}=\Omega $ for all $n$ except for $n=n_{0}$, where $\Omega
_{n_{0}}=\Omega _{def}$. The coordinate $x$ in Eq.\thinspace (\ref{Maxwell})
goes along the chain with the interatomic distance $a$, and the right hand
side of this equation is the microscopic polarization density. Parameter $%
\alpha $ is responsible for coupling between polarization and
electromagnetic waves. Eqs.\thinspace (\ref{exciton}, \ref{Maxwell}) present 
{\it microscopic} description of the transverse electromagnetic waves
propagating along the chain. These equations are subject to the boundary
conditions for the electromagnetic and polarization subsystems. We assume
that an incident and transmitted electromagnetic waves propagate in vacuum
so that the boundary conditions for Eq.\thinspace (\ref{Maxwell}) take the
usual form 
\begin{eqnarray}
E(0)=1+r; &&\qquad \frac{dE}{dx}{=ik(1-r);}  \label{Eboundary} \\
E(L)=t\exp {(ikL)}; &&\qquad \frac{dE}{dx}{=ikt\exp {(ikL),}}
\end{eqnarray}
where $k=\omega /c$ is a wave number of the electromagnetic wave in vacuum, $%
\mid t\mid ^{2}$ and $\mid r\mid ^{2}$ are transmission and reflection
coefficients, respectively, and $L$ is the length of the chain. The boundary
conditions for dipole excitations can be chosen in the general form 
\begin{equation}
\frac{P_{0}}{P_{1}}=\beta ;\qquad \frac{P_{N}}{P_{N}-1}=\gamma ,
\label{Pboundary}
\end{equation}
where $N=L/a$ is the number of sites in the chain, parameters $\beta $ and $%
\gamma $ describe different states of the ``surface'' of the chain. For
example, $\beta =0$, $\gamma =0$ correspond to the chain with fixed terminal
points. Another set of parameters, $\beta =1$, $\gamma =1,$ describes a
``relaxed surface'' where the forces on terminal sites are equal to zero. We
present results of calculations with these two choices of the boundary
conditions.

Our first goal is to convert the differential equation (\ref{Maxwell}) into
the discrete form. We can do this considering separately free propagation of
electromagnetic waves between sites and its scattering due to the
interaction with a dipole moment at the site. Let $E_{n}$ and $E_{n}^{\prime
}$ be the magnitude of the electromagnetic field and its derivative right
after scattering at the $n$th site. The electric field $E$ remains
continuous at a scattering site, while its derivative undergoes the jump,
which is equal to $-4\pi k^{2}P_{n}$. Finally, one can derive the system of
difference equations, that can be written with the use of the transfer
matrix, $T,$ in the form: 
\begin{equation}
v_{n+1}=T_{n}v_{n},  \label{EP}
\end{equation}
where we introduced the column vector, $v_{n},$ with components $P_{n}$, $%
P_{n+1}$, $E_{n}$, $D_{n}$ ($D_{n}=E_{n}^{\prime }/k$) and the transfer
matrix, $T_{n},$ that describes the propagation of the vector between
adjacent sites: 
\begin{equation}
T_{n}=\left( 
\begin{array}{cccc}
{0} & {1} & {0} & {0} \\ 
{-1} & {{-\displaystyle\frac{\Omega _{n}^{2}-\omega ^{2}}{\Phi }}} & %
\displaystyle{{\frac{\alpha }{4\pi }\cos {ka}}} & \displaystyle{{{\frac{%
\alpha }{4\pi }}\sin {ka}}} \\ 
{0} & {0} & {\cos {ka}} & {\sin {ka}} \\ 
{0} & {-4\pi k} & {-\sin {ka}} & {\cos {ka}}
\end{array}
\right) .  \label{T}
\end{equation}
The dynamical state of the system at the right end of the chain, which is
represented by the vector $v_{N}$, can be found from the initial state at
the left end, $v_{0},$ by means of the repetitive use of the transfer
matrix, $T$ : 
\begin{equation}
v_{N}=\prod_{1}^{N}T_{n}v_{0}.  \label{product}
\end{equation}
Since we consider the system with a single defect, all but one $T$-matrices
in Eq.\thinspace (\ref{product}) are the same. These matrices have the
parameter $\Omega _{n}$ (the only parameter, which distinguish the defect
site from the regular sites) equal to $\Omega $. For the matrix $T_{n_{0}}$,
which corresponds to the defect, $\Omega _{n}=\Omega _{def}$. The
eigenfrequencies of the pure system, i.e. polariton dispersion laws, can be
found by means of the eigenvalues of the $T$-matrix. There exist four
eigenvalues $\lambda $, which can be grouped in pairs with the product of
the members of each pair being equal to one. The eigenvalues can be found as
solutions of the following dispersion equation: 
\begin{equation}
\left( \lambda +\lambda ^{-1}-2\cos {ka}\right) \left( \lambda +\lambda
^{-1}+\frac{\Omega ^{2}-\omega ^{2}}{\Phi }\right) +4\pi k\sin (ka)=0.
\label{dispersion}
\end{equation}
In the band of propagating states, the solutions of Eq.\thinspace (\ref
{dispersion}) are complex valued numbers with their absolute values equal to
one. In this case the expression $\lambda +\lambda ^{-1}$ can be presented
in the form: $2\cos ({Qa),}$ where $Q$ is the Bloch wave number. With this
replacement the dispersion equation (\ref{dispersion}) takes the same form
as the equation obtained from the original Eqs.\thinspace (\ref{exciton}, 
\ref{Maxwell}) by means of Fourier transformation. It is important to
emphasize that Eq.\thinspace (\ref{dispersion}) takes into account the modes
of the electromagnetic field not only from the first Brillouin band but also
all short-wave components of the field. In this sense, our approach to the
problem is truly microscopic. In the band gap of the polariton spectrum, the
eigenvalues $\lambda $ become real valued and describe evanescent modes of
the system. Fig.\thinspace 1 presents the frequency dependence of the
absolute value of one of the eigenvalues. The band gap is clearly seen as a
region in which the absolute value of $\lambda $ is greater than 1.

We calculated the transmission coefficient of the electromagnetic waves
applying Eq.\thinspace (\ref{product}) to the vector $v_{0},$ with
components $\{P_{0},\beta P_{0},1+r,i(1-r)\}$, which describes the state of
electromagnetic waves and dipole subsystem at the left end of the chain in
accordance with boundary conditions (\ref{Eboundary}) and (\ref{Pboundary}).
The resulting state at the right end of the chain $v_{N}$ is to be matched
with the corresponding boundary conditions at $n=N$. We considered two kinds
of boundary conditions corresponding to fixed, $\beta =\gamma =0,$ and
relaxed, $\beta =\gamma =1,$ ends of the chain. For the numeric evaluation
we use the chain consisting of 30 atoms, with the defect placed at the $5$th
site. In order to check the computations, we calculated both transmission
and reflection coefficients and verified that the equality $\mid t\mid
^{2}+\mid r\mid ^{2}=1$ holds with sufficient accuracy.

The results of the calculations are presented in Figs.\thinspace 2 and 3,
which correspond to the fixed and relaxed boundary conditions respectively.
The parameter of the nearest neighbors interaction $\Phi $ was chosen to be
equal to $\Phi =\Omega ^{2}/3$, so that the maximum frequency of the
polarization waves is equal to $\sqrt{5/3}\Omega $. As one can see from
Fig.\thinspace 1, the polariton gap has a lower boundary at a slightly lower
frequency, $\simeq 1.24\Omega .$ It is caused by the negative dispersion of
the polariton waves assumed in the calculations. The frequency in all the
figures is normalized by $\Omega .$ The wavelength of the electromagnetic
waves in the region considered is much greater than the interatomic distance 
$a$, the product $ka$ is of order of $10^{-3}$, which corresponds to the
position of the exciton-polariton gap in real materials. Figs.\thinspace 2a
and 3a present the frequency dependence of the transmission in the pure
system for two types of boundary conditions. One can easily recognize the
boundary between the pass and stop bands in these figures. The transmission
exhibits a rich structure, corresponding to geometrical resonances due to
the finite size of the system in the pass band, and monotonically increases
with the frequency in the forbidden band. The increase of the transmission
is due to the frequency dependence of the penetration length $l=1/%
\mathop{\rm Im}%
q(\omega )$, where $q(\omega )$ is the imaginary wave number of polaritons
inside the gap.

All the other plots in Figs.\thinspace 2 and 3 show the transmission in a
system containing the defect for different values of the parameter $\Delta
=(\Omega _{def}^{2}-\Omega ^{2})/\Omega ^{2}$, which determines the strength
of the defect. One can see that the defect induces a resonant maximum at a
certain value of frequency, $\omega _{r},$ inside the forbidden gap. Though
resonance tunneling takes place for both kinds of boundary conditions, the
effect is much more prominent in the case of fixed ends. This fact is in
agreement with the overall greater transmission for the latter situation
than in the case of ``relaxed'' ends. The positions of the maxima was found
to be independent of the position of the defect in the chain as it should be
expected. We set the defect at different sites and found just a slight
modification of the shape of the maxima and their heights, but not the
positions.

The value of $\omega _{r}$ depends upon the strength of the defect, it moves
toward higher frequencies with increase of $\Delta $. This behavior is in
accord with the results of Refs.\thinspace \cite{Deych,Podolsky} regarding
the eigenfrequencies of local polariton states. For the model considered the
frequency of the local polariton is determined by an equation similar to
that obtained in Ref.\thinspace \cite{Deych}, 
\begin{equation}
1=\Delta \int_{-\pi }^{\pi }\displaystyle{\frac{\cos {(ka)}-\cos {x}}{\left(
k^{2}-1-\displaystyle\frac{2\Phi }{\Omega ^{2}}\cos {x}\right) \left( \cos {%
(ka)}-\cos {x}\right) -\displaystyle\frac{\alpha k}{2}\sin {(ak)}}}
\label{eigen}
\end{equation}
In Fig.\thinspace 4 we present the dependence of the resonance frequency, $%
\omega _{r},$ and the eigenfrequency of the local mode upon the defect
parameter, $\Delta $. One can see that these dependences are consistent with
each other. The deviation of $\omega _{r}$ from the eigenfrequency is
obviously due to the frequency dependence of the width of the resonance.

According to Eq.\thinspace (\ref{eigen}), local polaritons arise only for
positive $\Delta $. Indeed, when we change the sign, the resonance maximum
inside the stop-band disappears (Fig.\thinspace 5). At the same time, one
can identify in Fig.\thinspace 5a a new peak in the pass band, which arises
due to the defect. Though there are no new eigenstates in the region of the
continuous spectrum, the defect, nevertheless, causes resonance scattering
of propagating polaritons. This effect is known as quasilocal or resonance
``states,'' for example, in phonon physics \cite{Maradudin,Kosevich}, and
was discussed for polaritons by Hopfield \cite{Hopfield}. The defect-induced
maximum observed in the pass band in Fig.\thinspace 5a is caused by such
``quasilocal states.'' Surprisingly enough, is the absence of a quasilocal
resonance in Fig.\thinspace 5b. This represents the transmission in the case
of ``free'' surface. Instead, we observe a strong dip in the transmission
which was not present in the transmission of the pure chain. This situation
of antiresonance scattering is interesting but requires separate
consideration.

In conclusion, we have numerically shown that an electromagnetic wave with
frequency within the forbidden polariton band (``restrahlen region'') can
exhibit resonance tunneling via a microscopic defect, for example, an
impurity atom. The tunneling is due to the local polariton states associated
with the defect. Electromagnetic waves were treated in the paper
microscopically in the sense that we took into account the lattice structure
of the medium and all the modes of the electromagnetic field including those
with wave lengths shorter than $\pi /a$. The results of the calculations
showed that the short-wave components of the electromagnetic field indeed do
not contribute considerably to the effect, which is in agreement with the
assumption of Ref.\thinspace \cite{Deych}. The observed effects occurred for
long waves with wavelengths three orders of magnitude greater than the
interatomic distance.

Though we have considered the one-dimensional single-mode model, the main
result obtained in the paper -- the existence of resonance tunneling of
electromagnetic waves due to the local polariton states -- can be expanded
to more realistic situations. The one dimensional nature of the model
allowed for the microscopic treatment, which would not be feasible
otherwise. Once we have confirmed the assumption of Ref.\thinspace \cite
{Deych} regarding the role of shortwave component of electromagnetic waves,
one can treat the system in the framework of macroscopic methods and turn to
the consideration of more realistic models. Though the resonance tunneling
through a thin slab of a real 3-D material will have more complicated
properties, the essence of the effect will remain the same.

The most serious difference between our single-mode model of electromagnetic
waves and real situations is the absence of longitudinal modes in our model.
These modes could fill the gap between polariton branches and reduce the
lifetime of local polaritons. It is important to emphasize, however, that
unlike the above mentioned quasilocal states of propagating modes, the
transverse components of the local polariton remains localized. Therefore
the tunneling nature of electromagnetic wave propagation through the
restrahlen region is preserved even in the presence of the longitudinal
modes.

We wish to thank A.Z. Genack for reading and commenting on the manuscript
and J.L. Birman and V. Podolsky for useful discussions. This work was
supported by the NSF under grant No. DMR-9311605, by a CUNY collaborative
grant, and by a PSC-CUNY research award.

\pagebreak

\section*{Captions}

Fig.\thinspace 1. Frequency dependence of the absolute value of an
eigenvalue of the transfer matrix $T$. In the pass band, the absolute value
of $\lambda $ is equal to $1$, in the stop band it is greater than $1$.

Fig.\thinspace 2. Transmission through the chain with fixed ends. (a)
corresponds to the pure system, (b) and (c) describes the transmission
through the system with the defect for the different strength $\Delta $.

Fig.\thinspace 3. Same as in Fig.\thinspace 2 but for the chain with free
ends.

Fig.\thinspace 4. The solid line represents the relationship between
eigenfrequencies of the local polaritons and the defect parameter $\Delta $.
Two dashed lines show the positions of the resonance tunneling maxima for
different $\Delta $. The upper dashed line corresponds to the chain with
fixed ends, and the lower one presents the results for the chain with free
ends.

Fig.\thinspace 5. Quasilocal states in the pass band for the chain with (a)
fixed and (b) free ends.


\begin{references}
\bibitem{Rupasov}  V.I. Rupasov and M. Singh, Phys. Rev. A, {\bf 54}, 3614
(1996); Phys. Rev. A {\bf 56}, 898 (1997).

\bibitem{Deych}  Lev I. Deych and A.A. Lisyansky, {\it Local Polariton
States in Polar Crystals with Impurities, }Electronic Archive,
cond-mat/9706212 (1997); A.A. Lisyansky and Lev I. Deych, Bull. Amer. Phys.
Soc. {\bf 42, }203 (1997).

\bibitem{Podolsky}  V. Podolsky, Lev I. Deych and A.A. Lisyansky, Phys. Rev.
B, submitted.

\bibitem{Lifshitz}  I.M. Lifshitz, Nuovo Cim. (Suppl. A1), {\bf 3},591
(1956).

\bibitem{Maradudin}  A.A. Maradudin, E.W. Montroll, G.H. Weiss, and I.P.
Ipatova. {\it Theory of Lattice Dynamics in the Harmonic Approximation} 2nd
edition (Academic Press, New York) 1971.

\bibitem{Kosevich}  I.M. Lifshitz and A.M. Kosevich in {\it Lattice Dynamics}
(Benjamin, New York), p. 53 (1969).

\bibitem{Yablonovich}  E. Yablonovitch, T.J. Gmitter, R.D. Meade, A.M.
Rappe, K.D. Brommer and J.D. Joannopoulos, Phys. Rev. Lett., {\bf 67}, 3380
(1991).

\bibitem{Joannopulos}  R.D. Meade, K.D. Brommer, A.M. Rappe and J.D.
Joannopoulos, Phys. Rev. B, {\bf 44}, 13772 (1991).

\bibitem{Figotin}  A. Figotin and A. Klein, J. Stat. Phys., {\bf 86}, 165
(1997).

\bibitem{Hopfield}  J. J. Hopfield, Phys. Rev., {\bf 182}, 945 (1969).
\end{references}
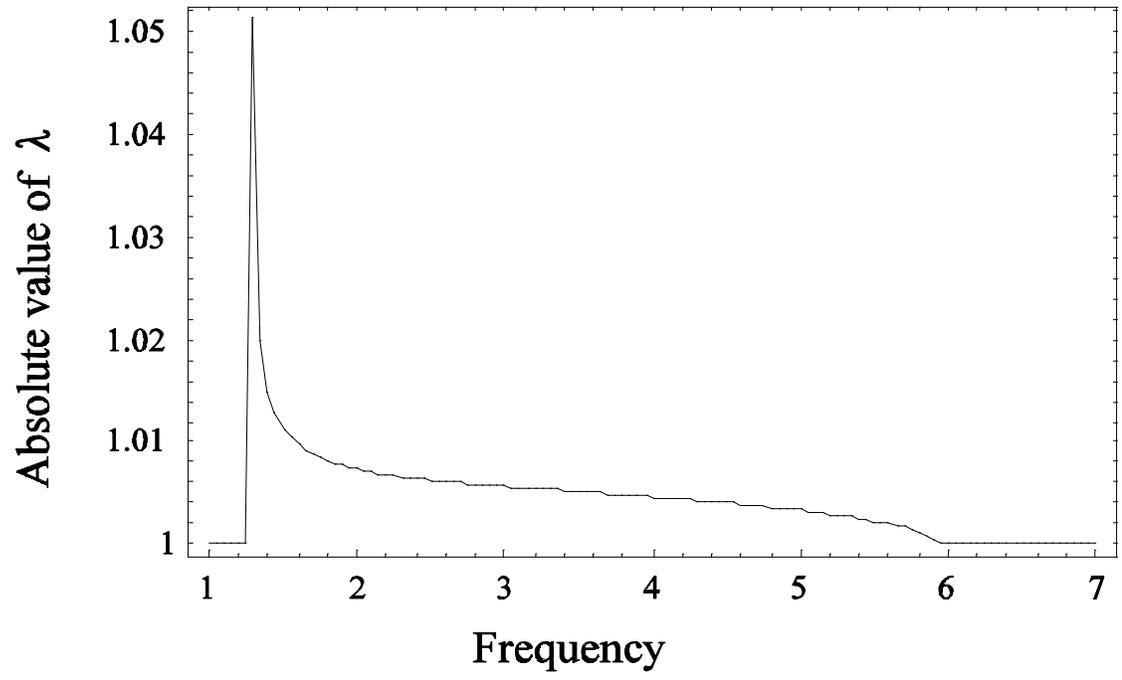
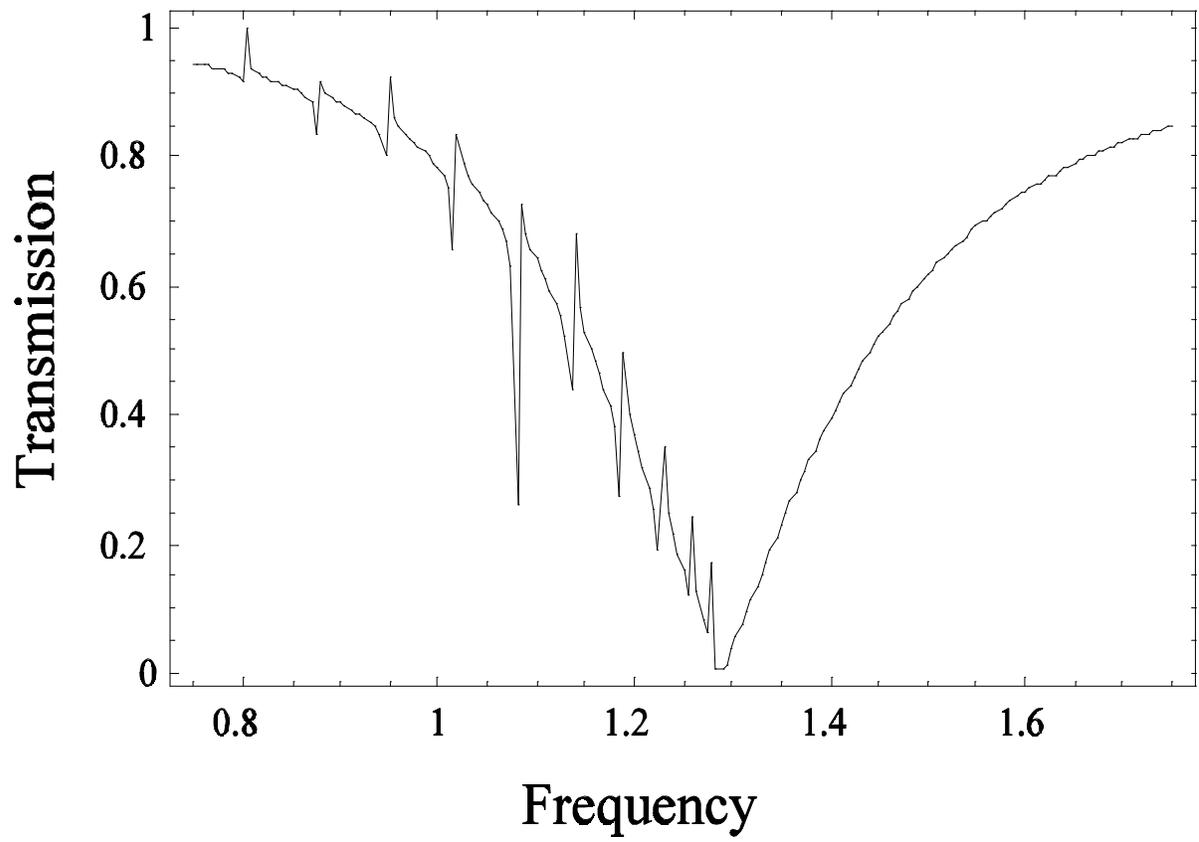
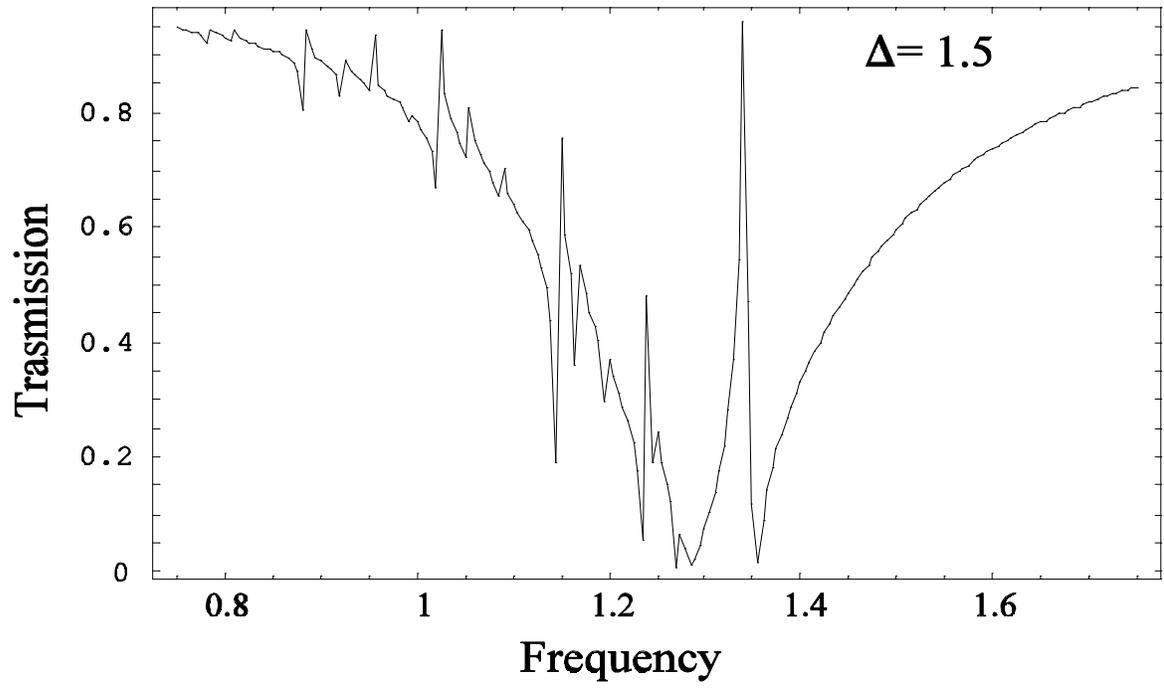
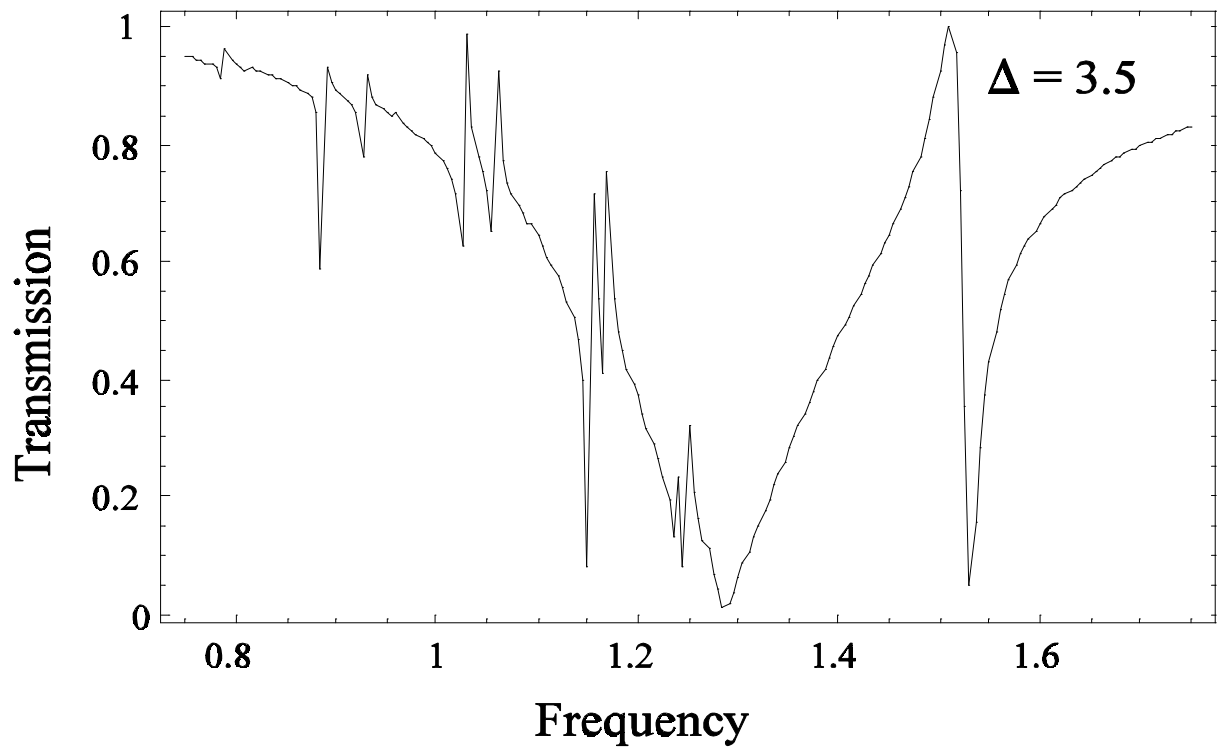
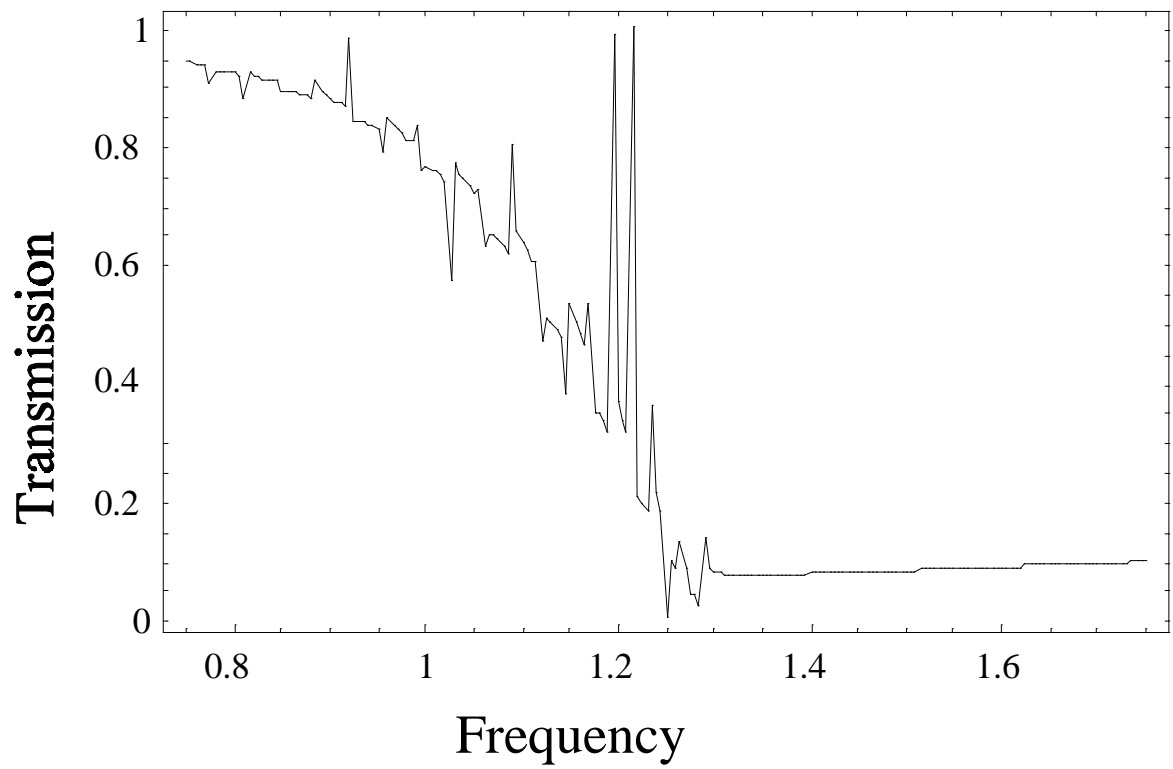
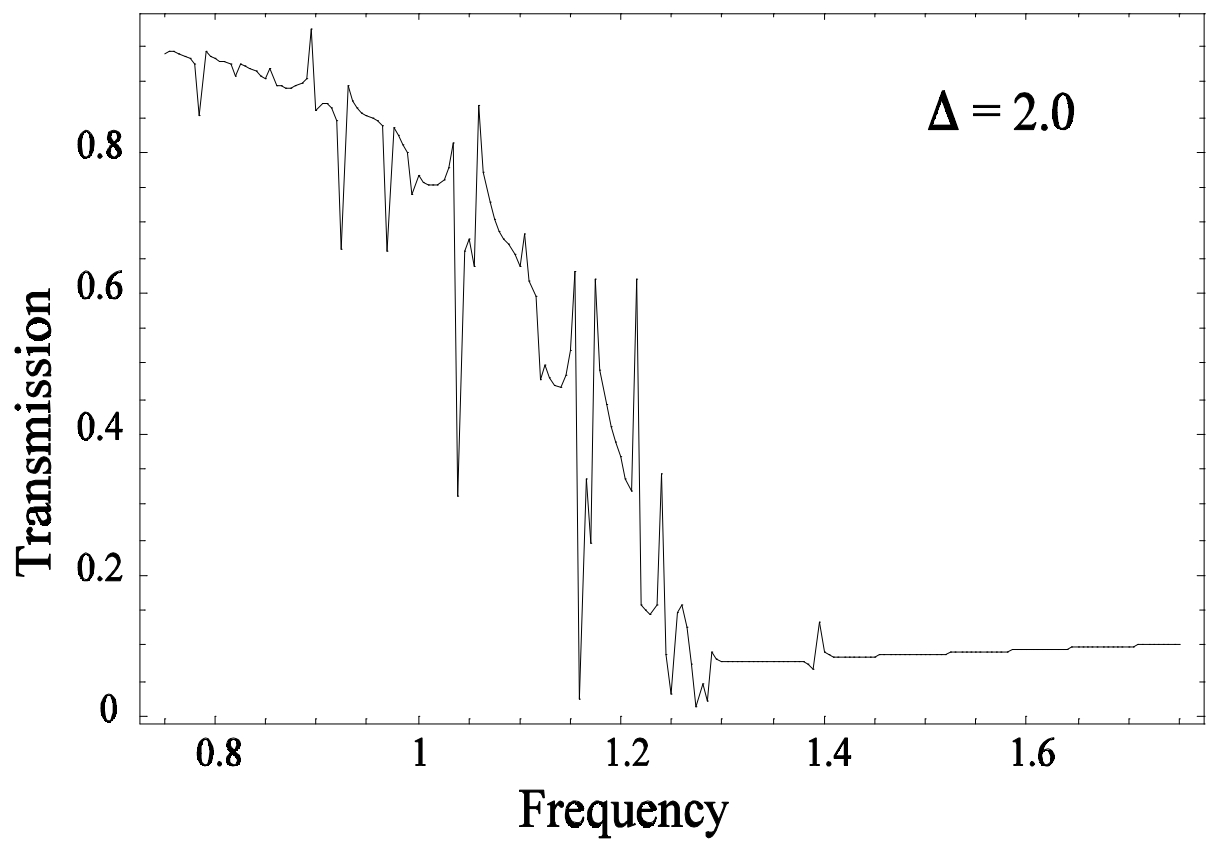
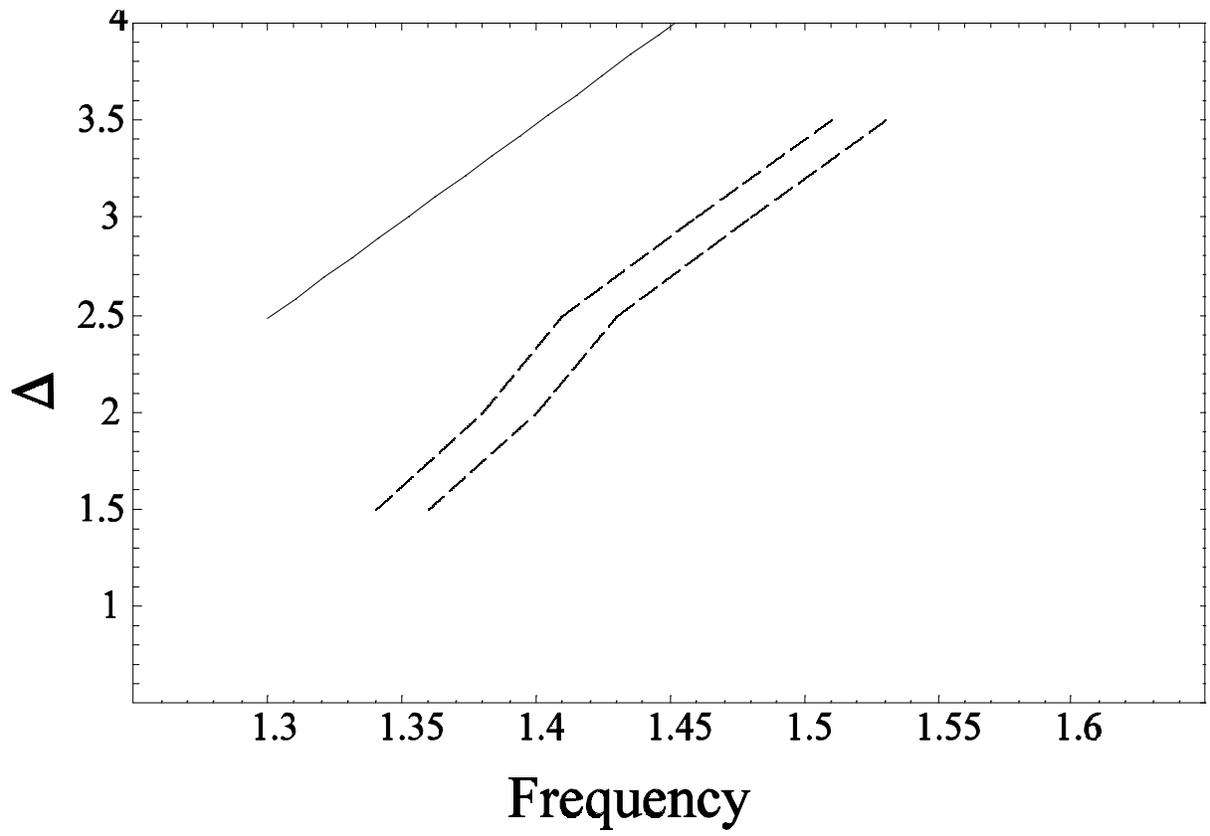
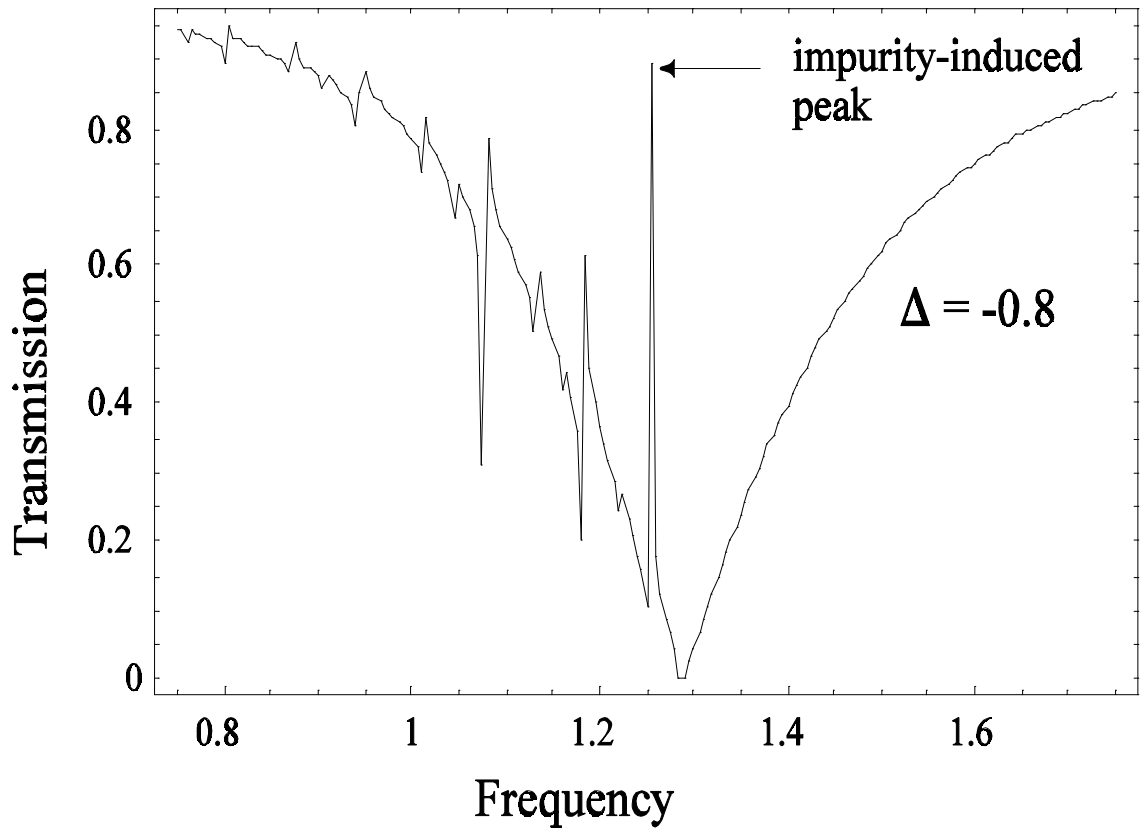
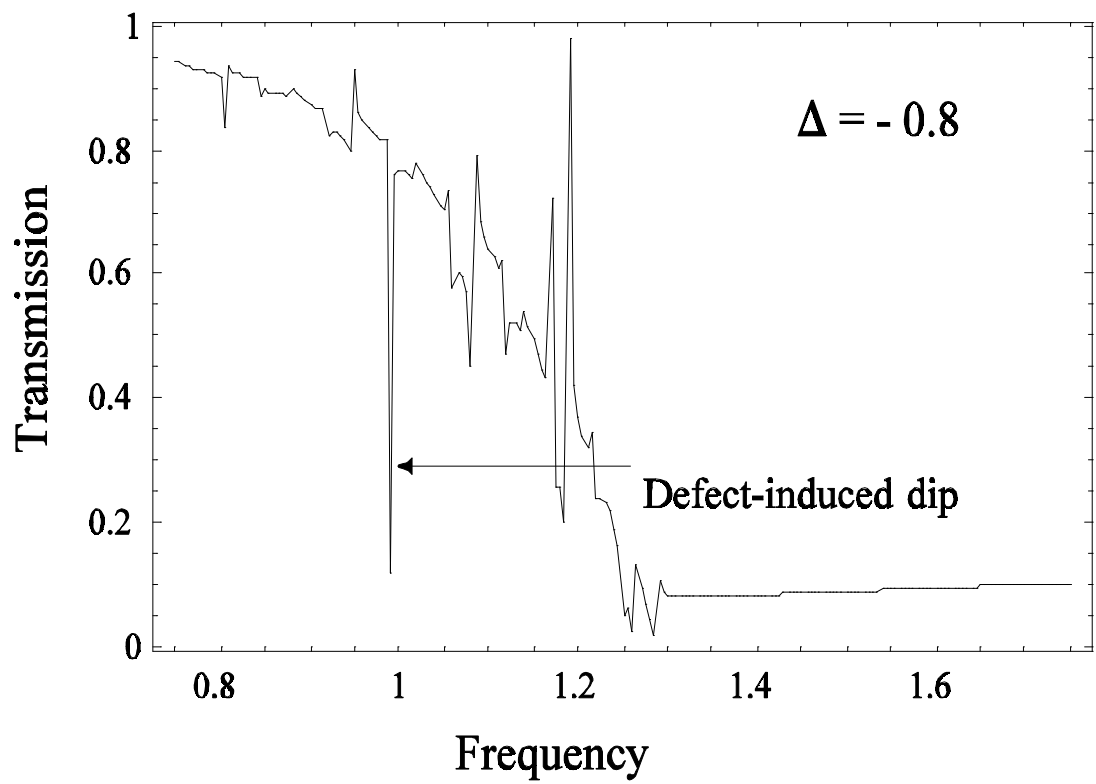
\end{document}